\documentclass{article}
\usepackage{spconf,amsmath,graphicx}
\usepackage{cite}
\usepackage{siunitx}
\usepackage{subfig}
\usepackage{url}
\usepackage{hyperref}
\usepackage{dcolumn}
\usepackage{amssymb}
\usepackage{amsbsy}
\usepackage{amsfonts}
\usepackage{booktabs}

\usepackage[linesnumbered,ruled,vlined,commentsnumbered]{algorithm2e}

\SetCommentSty{mycommfont}

\newcommand{\vect}[1]{\mbox{\boldmath $#1$}}
\newcommand{\abs}[1]{\left\lvert#1\right\rvert}

\DeclareMathOperator*{\argmax}{arg\,max}
\def\appendixautorefname~#1\null{~#1 \null}

\def\appendixautorefname~#1\null{~#1 \null}

\def\equationautorefname~#1\null{Eq.~(#1)\null}

\newcommand{\pdffigure}[3][width=0.7\linewidth]{
    \begin{figure}[tb]
    \begin{center}
    \IfFileExists{./#2.pdf}{
        \includegraphics[#1]{#2.pdf}
    }{
        \includegraphics[draft]{#2.pdf}
    }
    \end{center}
    \caption{#3}
    \label{fig:#2}
    \end{figure}
}
\title{Online End-to-End Neural Diarization with Speaker-Tracing Buffer}
\name{Yawen Xue$^1$, Shota Horiguchi$^1$, Yusuke Fujita$^1$, Shinji Watanabe$^2$, Paola Garc\' ia$^2$, Kenji Nagamatsu$^1$}
\address{$^1$Hitachi, Ltd. Research \& Development Group\\
  $^2$Center for Language and Speech Processing, Johns Hopkins University }
\begin{document}
%
\maketitle
\begin{abstract}
This paper proposes a novel online speaker diarization algorithm based on a fully supervised self-attention mechanism (SA-EEND). Online diarization inherently presents a speaker’s permutation problem due to the possibility to assign speaker regions incorrectly across the recording. To circumvent this inconsistency, we proposed a {\it speaker-tracing buffer mechanism} that selects several input frames representing the speaker permutation information from previous chunks and stores them in a buffer.
These buffered frames are stacked with the input frames in the current chunk and fed into a self-attention network. Our method ensures consistent diarization outputs across the buffer and the current chunk by checking the correlation between their corresponding outputs.
Additionally, we trained SA-EEND with variable chunk-sizes to mitigate the mismatch between training and inference introduced by the speaker-tracing buffer mechanism.
Experimental results, including online SA-EEND and variable chunk-size, achieved DERs of \SI{12.54}{\percent} for CALLHOME and \SI{20.77}{\percent} for CSJ with \SI{1.4}{\second} actual latency.
\end{abstract}
\section{Introduction}
With the recent advances in technology, audio-based human interaction systems are becoming quite popular. 
For them to work correctly, it is crucial to provide relevant information about the speakers and the speech transcription.
Speaker diarization --- which answers the question “who speaks when”--- is a crucial stage in the pipeline, since it can locate speaker turns and assign speech segments to speakers. 
Nowadays, speaker diarization has been widely studied in different scenarios; for example, meetings \cite{anguera2007acoustic, kang2020multimodal}, call-center telephone conversations \cite{callhome,martin2000nist}, and the home environment (CHiME-5, CHiME-6) \cite{barker2018fifth, kanda2018hitachi, watanabe2020chime}.

Currently, few speaker diarization systems can be applied in practical scenarios because most of them work well only under specific conditions such as long latency, no overlap, or low noise level \cite{von2019all,gmaciejewski2018characterizin}. 
An online speaker diarization system with low latency is still an open technical problem. 
Online speaker diarization outputs the diarization result as soon as the audio segment arrives, which means no future information is available when analyzing the current segment. 
In contrast, in an offline mode, the whole recording is processed so that all segments can be compared and clustered at the same time \cite{geiger2010gmm}.  

State-of-the-art speaker diarization systems mostly concentrate on integrating several components: voice activity detection, speaker change detection, feature representation, and clustering \cite{shum2013unsupervised,zhang2019fully}. 
Current research focuses primarily on the speaker model or speaker embeddings, such as Gaussian mixture models (GMM) \cite{geiger2010gmm, markov2008improved}, i-vector \cite{madikeri2015integrating, garcia2017speaker,zhu2016online}, d-vector \cite{wang2018speaker, wan2018generalized}, and x-vector \cite{snyder2018x, sell2014speaker}, and on a better clustering method such as agglomerative hierarchical clustering or spectral clustering \cite{snyder2018x, ning2006spectral, dimitriadis2017developing, patino2018low}. 
The issue with these methods is that they cannot directly minimize the diarization error because they are based on an unsupervised algorithm.
A supervised online speaker diarization method UIS-RNN \cite{zhang2019fully, fini2020supervised} was proposed while the method still assumes only one speaker in one segment (no overlapping).

To solve these issues, Fujita, et al. \cite{fujita2019end1,fujita2019end2, fujita2020end} proposed an end-to-end speaker diarization system (EEND).
Instead of applying several separate independent modules, EEND directly minimizes the diarization error by training a neural network using Permutation Invariant Training (PIT) with multi-speaker recordings. 
The experimental results show that the self-attention based end-to-end speaker diarization (SA-EEND) system \cite{fujita2019end2, fujita2020end} outperformed the state-of-the-art i-vector and x-vector clustering and long short-term memory (LSTM) \cite{fujita2019end1} based end-to-end method.     
Although SA-EEND has achieved significant improvement, it is only working in the offline condition which outputs speaker labels only after the whole recording is provided. 

This paper aims to extend offline SA-EEND to online speaker diarization. 
First of all, we investigate a straightforward online extension of SA-EEND by performing diarization independently for each chunked recording.
However, this straightforward online extension degrades the diarization error rate (DER) due to the speaker permutation inconsistency across the chunk, especially for short-length chunks.
To circumvent this inconsistency, our proposed method, called speaker-tracing buffer, selects several input frames representing the speaker permutation information from previous chunks and stores them in a buffer.
These buffered frames are stacked with the input frames in the current chunk and fed into the self-attention network together so that our method obtains consistent diarization outputs across the buffer and the current chunk by checking the correlation between the corresponding outputs.
Additionally, we also propose to train SA-EEND with variable chunk-sizes, which can mitigate the chunk size mismatch between training and inference due to the additional frames introduced by the above speaker-tracing buffer mechanism.
Lastly, we focus on the performance of the proposed method with respect to the original SA-EEND in an online situation by testing on the simulated dataset which was created using two speaker recordings with controlled overlap ratio, and two real datasets, CALLHOME and Corpus of Spontaneous Japanese (CSJ) datasets. 
In order to make our results reproducible, the code will be published online. 

\section{Related work}
Several online speaker diarization systems have already been developed along the last decade \cite{geiger2010gmm, markov2008improved, von2019all, zhu2016online, zhang2019fully, fini2020supervised}. 
Early online speaker diarization systems \cite{geiger2010gmm, markov2008improved} usually trained a Gaussian Mixture Model (GMM) from a huge amount of speech from different speakers to produce a universal background model.
When a speech region was assigned to a new speaker, maximum a posteriori adaptation was applied to adjust the GMM to these new speakers. 
Later on, the GMM approach was replaced by adapted i-vector or d-vector \cite{zhu2016online,zhang2019fully,fini2020supervised} which are referred to as speaker embeddings that represent individual information from each speaker.
The speaker embeddings are then compared and grouped using unsupervised or supervised clustering \cite{zhang2019fully, fini2020supervised}.
In \cite{von2019all}, an all-neural online approach that performed source separation, speaker counting, and diarization all together was proposed making it possible to optimize the entire online process. 
However, these systems are based on several separate modules and/or they can not directly minimize the diarization error.  

Recently, end-to-end neural networks have been successfully applied to various speech processing fields such as speech recognition, speech synthesis, and voice conversion.
Following this trend, Fujita et al., \cite{fujita2019end1,fujita2019end2} first proposed an end-to-end neural diarization (EEND) with several extensions, e.g., to deal with variable numbers of speakers in \cite{fujita2020neural, horiguchi2020end,lin2020optimal}.
While these extensions only consider an offline scenario, this paper focuses on extending the offline EEND method to an online scenario.  

\section{Analysis of online SA-EEND}
\subsection{SA-EEND}
In SA-EEND \cite{fujita2019end2}, the speaker diarization task is formulated as a probabilistic multi-label classification problem.
Given the $T$ length acoustic feature $X= \left(\vect{x}_t\in\mathbb{R}^{D}\mid t=1,\cdots,T\right)$, with a $D$-dimensional observation feature vector at time index $t$, SA-EEND predicts the corresponding speaker label sequence $\hat{Y}=\left({\hat{\vect{y}}}_t \mid t=1,\cdots,T\right)$. Here, speaker label ${\hat{\vect{y}}}_t = \left[\hat{y}_{t,s} \in \{0,1\} \mid s=1, \cdots, S\right]$ represents a joint activity for $S$ speakers at time $t$.
For example, $\hat{y}_{t,s}=\hat{y}_{t,s'}=1~\left(s \ne s'\right)$ means both $s$ and $s'$ spoke at time $t$. 
Thus, determining $\hat{Y}$ is the key to determine the speaker diarization information as follows:
\begin{equation}
    \hat{Y} = \mathrm{SA}\left(X\right)\in \left(0,1\right)^{S \times T},
    \label{eq:sa_eend}
\end{equation}
where $\mathrm{SA}(\cdot)$ is a multi-head self-attention based neural network. 

Note that the vanilla self-attention layers have to wait for all speech features in the entire recording to be processed in order to compute the output speaker labels.
Thus, this method causes very high latency determined by the length of the recording, and cannot be adequate for online/real-time speech interface.

\subsection{Chunk-wise SA-EEND for online inference}
This paper first investigates the use of SA-EEND as shown in \autoref{eq:sa_eend} for \textit{chunked} recordings with chunk size $\Delta$, as follows:
\begin{align}
    \label{eq:chunk_saeennd}
    \underbrace{\hat{Y}_{t_i + 1:t_i+\Delta}}_{\triangleq \hat{Y}_i} & = \mathrm{SA}(\underbrace{X_{t_i + 1:t_i+\Delta}}_{\triangleq X_i})\in \left(0,1\right)^{S \times \Delta}.
\end{align}
$i$ denotes a chunk index, and $t_{i=1} \triangleq 0$.
$X_i$ and $\hat{Y}_i$ denote subsequences of $X$ and $\hat{Y}$ at chunk $i$, respectively.
The latency can be suppressed by chunk size $\Delta$ instead of the entire recording length $T$.
We first investigate the influence of chunk size $\Delta$ in terms of the diarization performance.

\subsubsection{Model configuration and dataset}
The SA-EEND system was trained using simulated training/test sets for two speakers following the procedure in \cite{fujita2020end}. 
Here, four encoder blocks with 256 attention units containing four heads without residual connections were trained. 
The input features were 23-dimensional log-Mel-filterbanks concatenated with the previous seven frames and subsequent seven frames with a 25-ms frame length and 10-ms frame shift.
A subsampling factor of ten was applied afterwards. 
As a summary, a $(23\times 15)$-dimensional feature was inputted into the neural network every \SI{100}{\milli\second}. 

Two datasets were used for this analysis. 
The first one, a subset of CALLHOME \cite{callhome}, consists of actual two-speaker telephone conversations.
Following the steps in \cite{fujita2020end}, we split CALLHOME into two parts: 155 recordings for adaptation and 148 recordings for evaluation.
The overall overlap ratio (including the test set) is of \SI{13.0}{\percent}.
The average duration is \SI{72.1}{\second}. 
The second dataset is the Corpus of Spontaneous Japanese (CSJ) \cite{maekawa2003corpus} which consists of interviews, natural conversations, etc. 
We selected 54 recordings from this data with an overlap ratio of \SI{20.1}{\percent}. 
There are consistently two speakers in each recording with an average duration is \SI{767.0}{\second}. 

\subsubsection{Analysis results}
In this section, we analyzed the relationship between chunk size $\Delta$ in \autoref{eq:chunk_saeennd} and the DER.
The recordings to be analyzed were first divided equally according to predefined chunk size and then fed into a SA-EEND system.
These chunk-wise diarization results were then combined with the original order as the final diarization result of the whole recording.
We call it as recording-wise DER which was calculated on the entire recording.
When computing the DER in both overlapping and non-speech regions, a \SI{0.25}{\second} collar tolerance was used at the start and the end of each segment.

Note that this chunk-wise SA-EEND method does not guarantee that the permutation of speaker labels obtained across the chunk is the same due to the speaker permutation ambiguity underlying in the general speaker diarization problem.
Thus, the recording-wise DER would be degraded due to this across-chunk speaker inconsistency.
To measure this degradation, we also computed the oracle DER in each chunk separately (chunk-wise DER), which does not include the across-chunk speaker inconsistency error.

\begin{figure}[tb]
  \centering
  \begin{minipage}[t]{0.49\linewidth}
      \subfloat[CALLHOME]{
        \includegraphics[width=\linewidth]{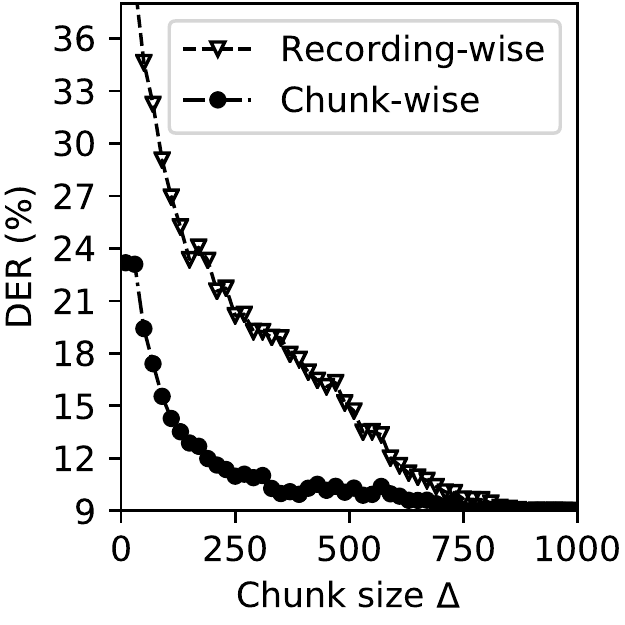}
        \label{fig:DER_Callhome_each_all}
      }
  \end{minipage}
  \hfill
  \begin{minipage}[t]{0.49\linewidth}
      \subfloat[CSJ]{
        \includegraphics[width=\linewidth]{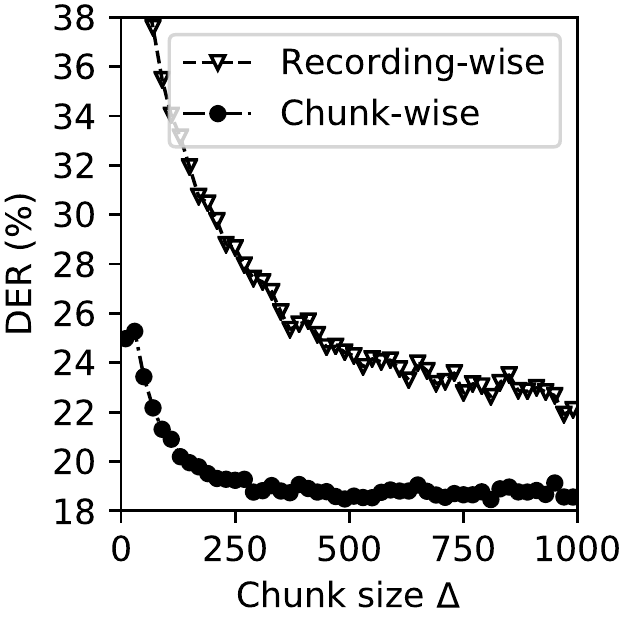}
        \label{fig:DER_CSJ_each_all}
      }
  \end{minipage}
    \caption{Recording-wise and oracle chunk-wise DER (\%).}
      \label{fig:chp2}
\end{figure}

The analytical results are shown in \autoref{fig:chp2} for the CALLHOME and CSJ datasets. 
In these figures, the x-axis represents a chunk size $\Delta$ during inference.
Here, one chunk unit corresponds to \SI{0.1}{\second}, which means the latency of the system (without considering the excution time) is \SI{1}{\second} when the chunk size is 10 (i.e., \SI{0.1}{\second} $\times$ 10 = \SI{1}{\second}).
The y-axis represents the final DER of the whole dataset. 
As shown in \autoref{fig:chp2}, the recording-wise DER decreased as the chunk size increased for both datasets.
When the chunk size was larger than 800, the recording-wise DER tended to converge for CALLHOME.
On the other hand, the oracle chunk-wise DER was much smaller and more stable than the recording-wise DER even when the chunk size was small, for both datasets.
This indicates that the main degradation of online chunk-wise SA-EEND comes from the across-chunk speaker permutation inconsistency.

\section{Speaker-tracing buffer}
In this section, we propose a method called speaker-tracing buffer (STB), that utilizes previous information as a clue to solve the across-chunk permutation issue. 

\begin{figure*}[t]
    \centering
	\includegraphics[width=0.9\linewidth]{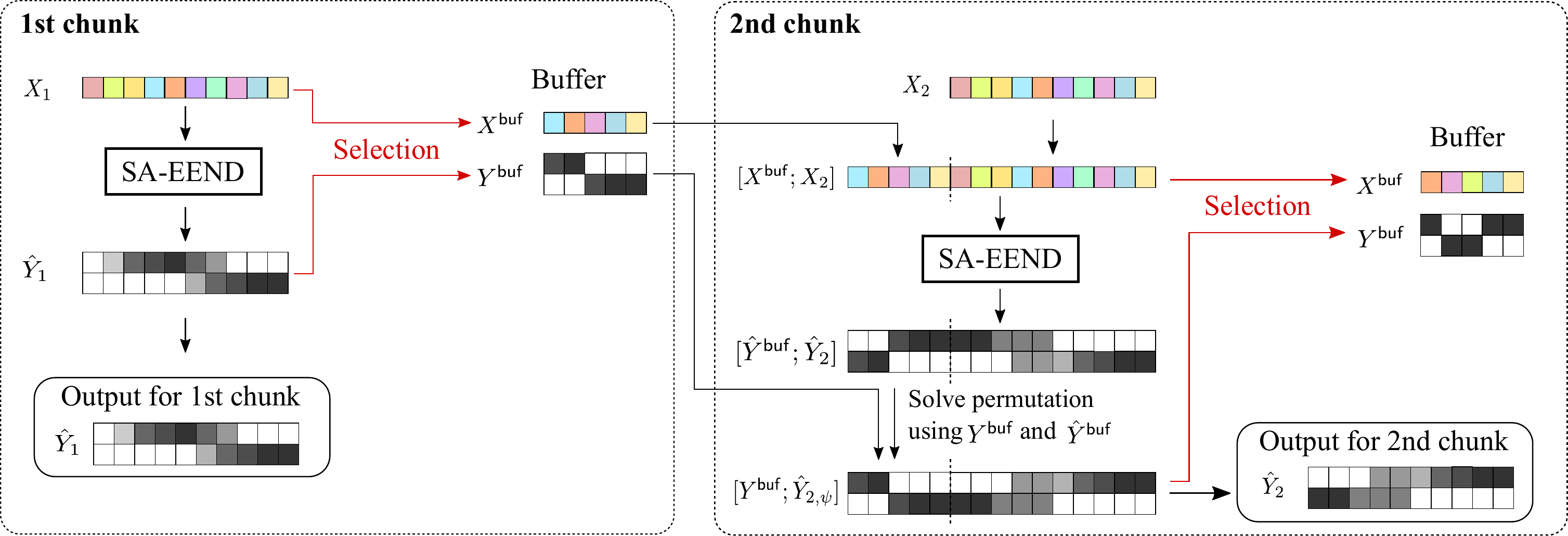}
 	\caption{Applying speaker-tracing buffer (STB) for SA-EEND.}
 	\label{fig:speak_tb}
\end{figure*}

\subsection{Speaker-tracing with buffer}
\begin{algorithm}[t]
    \SetAlgoLined
    \DontPrintSemicolon
    \caption{Online diarization using speaker-tracing buffer.}
    \label{alg:system}
    \SetAlgoVlined
    \SetKwInOut{Input}{Input}
    \SetKw{In}{in}
    \Input{
        {{$\{X_i\}_{i}$} \tcp*{Chunked acoustic subsequences}} \\
        {{$S$} \tcp*{\#speakers}}
        {{$L_\text{max}$} \tcp*{Buffer size}}
        {$\mathrm{SA}(\cdot)$ \tcp*{SA-EEND system}} 
    }
    \SetKwInOut{Output}{Output}
    \Output{{$\hat{Y}$} \tcp*{Diarization results}}
    \BlankLine
    $X^\mathsf{buf}\leftarrow\emptyset$, $Y^\mathsf{buf}\leftarrow\emptyset$ \tcp*{Initialize buffer}  
    \For{$i=1, \dots$}{
        $\left[\hat{Y}^\mathsf{buf}; \hat{Y}_i\right]\leftarrow \mathrm{SA}\left(\left[X^\mathsf{buf}; X_i\right]\right)$ \\
        \If{$Y^\mathsf{buf}\neq\emptyset$}{
            $\psi\leftarrow\argmax_{\phi\in \mathrm{perm}\left(S\right)}{\mathsf{CC}\left(Y^\mathsf{buf}, \hat{Y}^\mathsf{buf}_{\phi}\right)}$\\
            $\hat{Y}_i\leftarrow \hat{Y}_{i,\psi}$
        }
        $\hat{Y}\leftarrow\left[\hat{Y};\hat{Y}_i\right]$\\
        Update $X^\mathsf{buf}$ and $Y^\mathsf{buf}$ according to selection rules \tcp*{Sec. \ref{sec:selection_strategy}}
    }
\end{algorithm}
Let $L_\text{max}$ be the size of STB, and $X ^\mathsf{buf} \in \mathbb{R}^{D \times L}$ and $Y ^\mathsf{buf} \in \left(0,1\right)^{S \times L}~\left(0\leq L\leq L_\text{max}\right)$ be the $L$-length acoustic feature and the corresponding SA-EEND outputs stored in STB, respectively, which contain the speaker-tracing information.
At the initial stage, $X^\mathsf{buf}$ and $Y^\mathsf{buf}$ are empty. 
Our online diarization is performed by referring and updating this STB, as shown in \autoref{alg:system}.
The input of the SA-EEND system is the concatenation of acoustic feature subsequence $X_i\in \mathbb{R}^{D\times \Delta}$ at current chunk $i$ and the acoustic features in buffer $X^\mathsf{buf}$, i.e., $\left[X^\mathsf{buf}; X_i\right] \in \mathbb{R}^{D\times \left(L + \Delta\right)}$.
The corresponding output of SA-EEND is $\left[\hat{Y}^\mathsf{buf}; \hat{Y}_i\right] \in \left(0,1\right)^{S \times \left(L + \Delta\right)}$. 
If $Y^\mathsf{buf}$ is not empty, the correlation coefficient $\mathsf{CC}\left(\cdot, \cdot\right)$ between $Y^\mathsf{buf}$ and the current buffer output $\hat{Y}^\mathsf{buf}_{\phi}$ at speaker permutation output $\phi$ is calculated as 

\begin{multline*}
    \mathsf{CC}\left(Y^\mathsf{buf}, \hat{Y}^\mathsf{buf}_{\phi}\right)=
    \\
    {\frac{\sum_{s=1}^{S} \sum_{l=1}^{L} \left(y^\mathsf{buf}_{s,l} - \overline{y^\mathsf{buf}}\right)\left(\hat{y}^\mathsf{buf}_{\phi_s,l} - \overline{\hat{y}^\mathsf{buf}_{\phi}}\right)}
{\sqrt{\sum_{s=1}^{S} \sum_{l=1}^{L}\left(y^\mathsf{buf}_{s,l} - \overline{y^\mathsf{buf}}\right)^2}{\sqrt{\sum_{s=1}^{S} \sum_{l=1}^{L}\left(\hat{y}^\mathsf{buf}_{\phi_s,l} - \overline{\hat{y}^\mathsf{buf}_{\phi}}\right)^2}}}}, \label{eq:cc}
\end{multline*}
where
\begin{equation}   
\overline{y^\mathsf{buf}}=\frac{\sum_{s=1}^S\sum_{l=1}^Ly_{s,l}^\mathsf{buf}}{SL},\quad
\overline{\hat{y}_\phi^\mathsf{buf}}=\frac{\sum_{s=1}^S\sum_{l=1}^L\hat{y}_{\phi_s,l}^\mathsf{buf}}{SL}.
\end{equation}

Permutation $\psi$ with the largest correlation coefficient is chosen as follows:
\begin{equation}
    \psi = \argmax_{\phi\in \mathrm{perm}\left(S\right)}{\mathsf{CC}\left(Y^\mathsf{buf}, \hat{Y}^\mathsf{buf}_{\phi}\right)},
\end{equation}
where $\mathrm{perm}(S)$ generates all permutations according to the number of speakers $S$.
The corresponding buffer output $\hat{Y}_{i, \psi}^\mathsf{buf}$ is chosen as the final output $\hat{Y}_i$ of chunk $i$, which can maintain a consistent speaker permutation across the chunk.
The obtained output $\hat{Y}_i$ is stacked with the previously estimated output to form the whole recording's output $\hat{Y}$ in the end.
An example of applying the STB to SA-EEND in the first two chunks is shown in \autoref{fig:speak_tb}, where $\Delta$ is equal to 10, the buffer size $L_\text{max}$ is 5, and the speaker number $S$ is 2.

Speaker-tracing buffer $\left(X^\mathsf{buf}; Y^\mathsf{buf}\right)$ for the next chunk $i+1$ is selected from $\left[Y^\mathsf{buf}; \hat{Y}_i\right]$ and $\left[X^\mathsf{buf}; X_i\right]$ in the current chunk.
We consider four selection strategies, as explained in the next section.

\subsection{Selection strategy for speaker-tracing buffer}\label{sec:selection_strategy}
If chunk size $\Delta$ is not larger than the predefined buffer size $L_\text{max}$, we can simply store all the features in the buffer until the number of stored features reaches the buffer size.
Once the number of accumulated features becomes larger than the buffer size $L_\text{max}$, we have to select and store informative features that contain the speaker permutation information from $\left[X^\mathsf{buf};X_i\right]$ and $\left[Y^\mathsf{buf};\hat{Y}_i\right]$.
In this section, four selection rules for updating the buffer are listed.
Here, we assume that the number of speakers $S$ is 2.

\begin{itemize}
    \item \textbf{First-in-first-out}. The buffer is managed in a first-in-first-out manner to store the most recent $L_\text{max}$ features.
    \item \textbf{Uniform sampling}. 
    $L_\text{max}$ acoustic features from $\left[X^\mathsf{buf};X_i\right]$ and the corresponding diarization results from $\left[Y^\mathsf{buf};\hat{Y}_i\right]$ are randomly extracted based on the uniform distribution.
    \item \textbf{Deterministic selection} using the absolute difference of probabilities of speakers, as
        \begin{align}
            \delta_{m} = \abs{y_{m,1} - y_{m,2}},
        \label{eq:AF}
        \end{align}
    where $y_{1,m}$, $y_{2,m}$ are the probabilities of the first and second speakers at time index $m$.
    The maximum value of $\delta_m~\left(=1\right)$ is realized in either case of $y_{m,1} = 1, y_{m,2} = 0$ or $y_{m,1} = 0, y_{m,2} = 1$.
    This means that we try to find dominant active-speaker frames.
    Top $L_\text{max}$ samples with the highest $\delta_{m}$ are selected from $\left[X^\mathsf{buf};X_i\right]$ and $\left[Y^\mathsf{buf};\hat{Y}_i\right]$
    \item \textbf{Weighted sampling}:
    This is a combination of the uniform sampling and deterministic selection.
    We randomly select $L_\text{max}$ features but the probability of selecting $m$-th feature is proportional to $\delta_{m}$in \autoref{eq:AF}.
\end{itemize}

\subsection{Efficient training scheme for SA-EEND}
The additional frames introduced by the speaker-tracing buffer mechanism cause the length of the input chunks to get larger overtime at the inference stage. 
However, the model trained in \cite{fujita2019end2} used a fixed chunk size.
This will originate a mismatch between training and evaluation, which would degrade the performance of this system \cite{jung2019short, kanagasundaram2019study}. 

Since self-attention modeling does not depend on the input length, we propose a variable chunk size training (VCT) scheme to mitigate the chunk size mismatch issue. 
First, we split each recording into chunks with size $\gamma$, randomly sampled from $\left\{50,51,\dots,500\right\}$. 
When we created a minibatch, it contained variable-length sequences due to the variable chunk size, and we used a padding technique to compensate for the different lengths.
VCT scheme is applied to both training and adaptation stages. 
Due to the padding technique, the training efficiency was marginally degraded compared with fixed chunk-size training.
%

\section{Experimental results}
\subsection{Effect of selection strategy}
We analyzed the effect of the speaker tracing buffer (STB) and the selection strategy in the section.
For the effect of selection strategy, we used the same chunk size $\Delta=10$ and several buffer sizes $L_\text{max}$ varied from 10 to 1000 in \autoref{tab:chp4.1-2}.
The model used here is the fixed-length training model with four encoder blocks and four heads. 

Note that online diarization without STB with $\Delta=10$ showed \SI{38.29}{\percent} and \SI{44.57}{\percent} DERs on CALLHOME and CSJ, respectively.
Comparing online diarization without STB and with STB, applying the STB improved the performance of online SA-EEND regardless of which selection strategy was used.
As for the strategies, weighted sampling performed best for both datasets in most cases when $L_\text{max}$ was large.
Therefore, we considered weighted sampling as a selection strategy for future analysis. 

\begin{table}[t]
    \centering
    \caption{DERs (\%) of online diarization using STB with four buffer selection strategies. We varied buffer size $L_\text{max}$, but fixed chunk size $\Delta=10$. Note that online diarization without STB with $\Delta=10$ showed \SI{38.29}{\percent} and \SI{44.57}{\percent} DERs on CALLHOME and CSJ, respectively.}
    \label{tab:chp4.1-2}
    \subfloat[CALLHOME]{
        \resizebox{\linewidth}{!}{
        \begin{tabular}{@{}lcccccc@{}}
            \toprule
            &\multicolumn{6}{c}{$L_\text{max}$}\\\cmidrule(l){2-7}
            System&10&50&100&200&500&1000\\\midrule
            First-in-first-out & 48.71  & 29.49  & 18.05  & 14.09  & \bf12.80 & 12.66\\
            Uniform sampling &  45.03&22.38&16.28&13.95&13.05 &\bf12.65 \\
            Deterministic selection &37.56&23.23&17.11&14.47&\bf12.80 &12.66 \\
            Weighted sampling&42.23&\bf20.10&\bf15.47&\bf13.26&12.84&12.66\\
            \bottomrule
        \end{tabular}
        }
    }\\
    \subfloat[CSJ]{
        \resizebox{\linewidth}{!}{
        \begin{tabular}{@{}lcccccc@{}}
            \toprule
            &\multicolumn{6}{c}{$L_\text{max}$}\\\cmidrule(l){2-7}
            System&10&50&100&200&500&1000\\\midrule
            First-in-first-out & 51.32  & 44.08 & 37.22 & 26.21 & 22.02 & \bf20.45\\
            Uniform sampling &\bf39.24&31.06&26.38&24.99&24.51&20.59 \\
            Deterministic selection &45.70&\bf29.89&27.06&25.32&24.70&24.13\\
            Weighted sampling&49.87&30.11&\bf25.44&\bf22.69&\bf21.64&21.62\\
            \bottomrule
        \end{tabular}
        }
    }
\end{table}

\subsection{Effect of buffer and chunk size}
\begin{figure}[tb]
  \centering
  \begin{minipage}[t]{0.49\linewidth}
      \subfloat[CALLHOME]{
        \includegraphics[width=\linewidth]{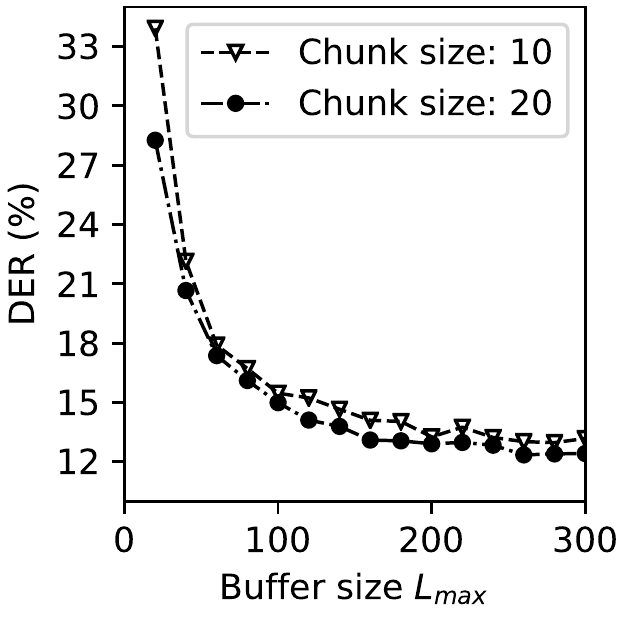}}
  \end{minipage}
  \hfill
  \begin{minipage}[t]{0.49\linewidth}
      \subfloat[CSJ]{
        \includegraphics[width=\linewidth]{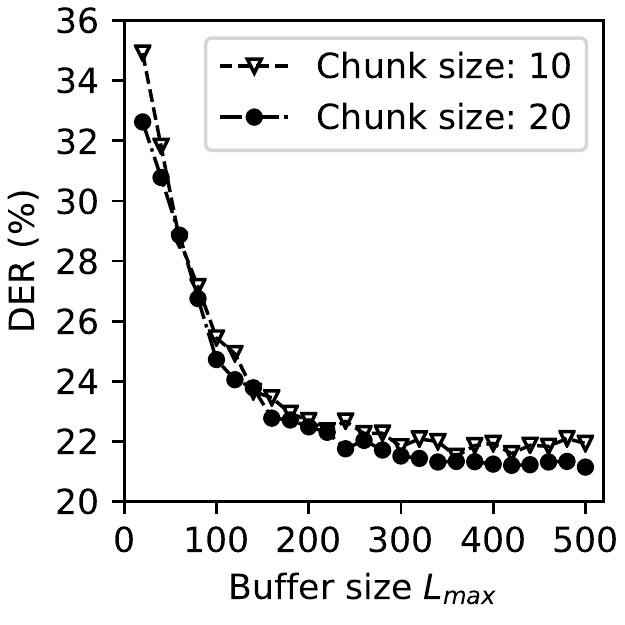}
      }
  \end{minipage}
  \caption{Relationship among DER, chunk size and buffer size.}
  \label{fig:chp4}
\end{figure}

Next, we analyzed the effect of the buffer and the chunk size.
The DER results for the CALLHOME and CSJ when applying the weighted sampling selection strategy are shown in \autoref{fig:chp4}.
Chunk sizes $\Delta$ were 10 and 20 with the latency time of \SI{1}{\second} and \SI{2}{\second} respectively. 
Regarding the chunk size in \autoref{fig:chp4}, all DERs from the large chunk size $\Delta=20$ are better than those from the small chunk size $\Delta=10$ even if the buffer size is the same. 
As for the buffer size, DER decreased as buffer size increased. 
These results were in line with our assumption that a large input size would lead to a better result. 
The model used here was the fixed-length training model with four encoder blocks and four heads. 

\subsection{Real-time factor}
Real-time factor (RTF) was calculated as the ratio between the summation of the execution time of every chunk to a recording duration
It measures the speech decoding speed and expresses the time performance of the proposed system.
To avoid the unequal size of buffers in the first several chunks, we first filled the buffer with dummy values and then calculated the RTF.
Our experiment was conducted on an Intel\textsuperscript{\textregistered} Xeon\textsuperscript{\textregistered} CPU E52697A v2 @ 2.60GHz using one thread. 
RTFs are equal to 0.40, 1.07 when the chunk size $\Delta=10$ and the buffer size $L_\text{max}=500,1000$.
This indicates that the proposed method is acceptable for online applications when buffer size is smaller than 1000 (\SI{100}{\second}).
So, the actual latency time for online SA-EEND is \SI{1.4}{\second} when $\Delta=10$ and $L_{\text{max}}=500$. 

\subsection{Comparison with other methods}

\begin{table*}[hbt!]
    \caption{DERs (\%) on simulated mixtures and real datasets. $\rho$ denotes the overlap ratio of each simulated dataset. \textit{Note that all results include the overlapping regions without oracle speech activity detection}.}
    \label{tab:chp4.4-1}
    \centering
    \resizebox{0.8\linewidth}{!}{
    \begin{tabular}{@{}lccccc@{}} 
        \toprule
        & \multicolumn{3}{c}{Simulated} & \multicolumn{2}{c}{Real} \\
        \cmidrule(lr){2-4}\cmidrule(l){5-6}
        System & $\rho=\SI{34.4}{\percent}$ & $\SI{27.2}{\percent}$ &$\SI{19.5}{\percent}$ & CALLHOME & CSJ  \\ \midrule
        Offline i-vector   & 33.73 & 30.93 & 25.96 & 12.10 & 27.99  \\
        Offline x-vector  & 28.77 & 24.46 & 17.78 & 11.53 & 22.96  \\
        Offline SA-EEND  ($\gamma=500$) & {\;\:}{\bf 4.56} & {\;\:}{\bf 4.50} & {\;\:}{\bf 3.85} & {\;\:}{\bf9.54} & {\bf 20.48} \\
        \midrule
        Online x-vector ($\Delta=15$)& 36.94  & 34.94  &33.19   & 26.90 & 25.45 \\ 
        Online SA-EEND ($\Delta=10$) & 33.18 & 37.31 & 41.41 & 38.29 & 44.57 \\
        Online SA-EEND w/ STB ($\Delta=10,L_\text{max}=500$)&  {\;\:}7.91 &  {\;\:}7.31 &  {\;\:}6.91 &  12.84 &  21.64 \\
        Online SA-EEND w/ STB ($\Delta=5,L_\text{max}=500$)& {\;\:}7.87 & {\;\:}7.48 & {\;\:}7.15 &  13.08 & 22.54 \\
        Online SA-EEND w/ STB and VCT ($\Delta=10,L_\text{max}=500$) & {\;\:}\bf 7.41    &  {\;\:}\bf 6.98 & {\;\:}\bf 6.27 & \bf 12.54   & \bf 20.77 \\
        Online SA-EEND w/ STB and VCT ($\Delta=5,L_\text{max}=500$) & {\;\:}7.79 & {\;\:}7.53 & {\;\:}6.88 & 12.66 & 21.62\\
        \bottomrule
    \end{tabular}
    }
\end{table*}
\begin{table*}[hbt!]
    \caption{DERs (\%) on real datasets with \SI{30}{s} of calibration period.}
    \label{tab:chp4.4-2}
    \centering
    \resizebox{0.8\linewidth}{!}{
    \begin{tabular}{@{}lcccccc@{}} 
        \toprule
        & \multicolumn{3}{c}{CALLHOME} & \multicolumn{3}{c}{CSJ}\\
        \cmidrule(lr){2-4}\cmidrule(l){5-7}
        System & Within \SI{30}{\second}   & After \SI{30}{\second} & All &  Within \SI{30}{\second}   & After \SI{30}{\second} & All \\ \midrule
       Offline SA-EEND 
       & {\;\:}9.38 & {\;\:}9.53 & {\;\:}9.53 & 23.43 & 20.23   & 20.48 \\
       Online SA-EEND w/ STB 
       & 15.91 & 10.05 & 12.84 & 25.37& 21.30 & 21.64 \\
       Online SA-EEND w/ STB and VCT  
       & 14.89 & 10.58& 12.54& 23.52& 20.49  & 20.77 \\
        \bottomrule
    \end{tabular}
    }
\end{table*}
For a comparison with other methods, we evaluated our proposed methods using two real datasets (CALLHOME and CSJ), and three simulated datasets which are shown in \autoref{tab:chp4.4-1}. 
The simulated datasets were created by using two speaker segments.
The background noise and room impulse response come from MUSAN corpus \cite{snyder2015musan} and Simulated Room Response corpus \cite{ko2017study} following the procedure in \cite{fujita2019end2}. 
Three kinds of simulated datasets were created with overlap ratios equal to $\rho=\SI{34.4}{\percent}$, \SI{27.2}{\percent}, and \SI{19.5}{\percent}, respectively. 

Note that we included errors in overlapping speech segments and speech-activity-detection-related errors in contrast with most works, e.g., the Kaldi CALLHOME diarization recipe, that did not evaluate such errors.

For the offline i-vector and x-vector method, we used version 1 and 2 (v1 and v2) from Kaldi CALLHOME diarization recipe
\cite{snyder2017deep,snyder2018x,povey2011kaldi}.
These are offline methods that employ probabilistic linear discriminant analysis \cite{ioffe2006probabilistic} along with agglomerative-cluster, a TDNN-based speech activity detection \cite{peddinti2015jhu} and oracle number of speakers.
Offline SA-EEND refers to the method that uses the entire recording as one chunk. The system in \cite{fujita2020end} which achieved the best performance is applied here, not only for the offline SA-EEND but also for the online SA-EEND w/ STB. 
For the online SA-EEND, the chunk size is $\Delta=10$ without applying STB. 
The proposed method applied the weighted sampling based STB as shown in \autoref{sec:selection_strategy} to the SA-EEND.

For the online x-vector, the speech segments were divided into subsequent \SI{1.5}{\second} chunks ($\Delta = 15$). 
Then, the system decided whether the entire chunk was speech or silence based on the output of the energy VAD for real datasets and oracle VAD for simulated datasets.
If the percentage of voiced frames of the entire chunk was fewer than \SI{20}{\percent}, it was considered as silence, and the process was skipped.
If it was a voiced chunk, we extracted an x-vector and assigned it to the first cluster until a dissimilar x-vector arrives according to the probabilistic linear discriminant analysis (PLDA) score.
Here, we applied a suitable threshold of 0.1 as the dissimilar criterion after scanning thresholds from 0.2 to -0.2 with a step of 0.1. 
Once two clusters exist we computed the PLDA score between the new segment and the two clusters.
Finally, we assigned an x-vector to the nearest cluster.

As shown in \autoref{tab:chp4.4-1}, among these online systems including the system based on x-vector, online SA-EEND with STB and VCT achieved the best result.
The proposed online SA-EEND performed even better than the offline i-vector and x-vector based methods on the CSJ dataset.

The online SA-EEND w/ STB and VCT increased the DER by about \SI{3}{\percent} when compared with the offline SA-EEND system for CALLHOME and Simulated dataset when buffer size is 500 and the chunk size is 10.
It is also shown that the online SA-EEND with STB and VCT almost achieved the offline performance by \SI{20.77}{\percent} DER as the DER of the offline SA-EEND is \SI{20.48}{\percent}.
As the average duration of the recordings in CSJ (\SI{767.0}{\second}) is much longer than CALLHOME (\SI{74}{\second}) and Simulated (\SI{87.6}{\second}).
It indicates that online SA-EEND with STB can achieve better results for long recordings.

In order to explore the increase of DER, we broke down the DER with a calibration period of \SI{30}{\second} as described in \autoref{tab:chp4.4-2}.
We can observe that within the 30s, the model trained with VCT performs better than with a fixed length. 
But after the 30s, the model trained with fixed length can achieve better results. 
Both methods show comparable results with offline SA-EEND after the 30s which also explains the reason why STB is much more suitable for long recordings.  


\section{Conclusion}
In this paper, we proposed a speaker-tracing buffer to memorize the permutation information of the previous chunk which enables the pre-trained offline SA-EEND system directly work online.
In addition, the variable chunk size training scheme was proposed to handle the variable input length using speaker tracing buffer.  
The latency time can be reduced to \SI{500}{\milli\second} with comparable diarization performance. 
Future work will be focused on a variable number of speakers as the current method is limited to the two-speaker case. 
In addition, the combination of fixed-length training and the variable training scheme will be considered.

\clearpage
\bibliographystyle{IEEEbib}
\bibliography{strings,refs}

\end{document}